\documentstyle[pre,aps,multicol,epsfig,,eqsecnum,color]{revtex}

\begin{document}
\draft     

\title{ An individual based model with global competition interaction: 
fluctuation effects in pattern formation}

\author{E. Brigatti $^{\star\pm}$, V. Schw\"ammle$^{\dag}$ and Minos A. Neto$^{\star}$}  
  
\address{$\dag$Centro Brasileiro de Pesquisas F\'{\i}sicas, Rua Dr. Xavier 
  Sigaud 150, 22290-180, Rio de Janeiro, RJ, Brazil} 
\address{$\star$Instituto de F\'{\i}sica, Universidade Federal Fluminense, 
  Campus da Praia Vermelha, 24210-340, Niter\'oi, RJ, Brazil}
\address{$\pm$e-mail address: edgardo@if.uff.br}

\date{\today}
\maketitle
\widetext
  
\begin{abstract}
 
We present some numerical results obtained 
from a simple individual based model that describes 
clustering of organisms caused by competition. 
Our aim is to show that, even when a deterministic description  
developed for continuum models predicts no pattern formation, 
an individual based model  
displays well defined patterns, as a consequence of 
fluctuation effects caused by the discrete nature of the 
interacting agents.

\end{abstract}
  
   

\begin{multicols}{2}

\section{Introduction}

Birth and death processes are two of the most relevant characteristics of 
the dynamics of biological populations and
can be responsible for the emergence of stable spatial patterns \cite{reviewpattern}.
In fact, the intrinsic asymmetry in the nature of birth and death processes
can enhance small initial differences in the 
spatial population density and lead to the formation of structures \cite{natureth,natureth2,natureth3}.
These clusters are resistant to some levels
of diffusion and emerge as soon as the birth of new
individuals outcompetes their movements.
For this reason, simplified models combining birth and death 
processes with Brownian movement are able to describe aggregation of 
individuals \cite{nature}.

Another central ingredient, present in ecological systems,
that can cause the generation of spatial structures, is  
the competition for resources \cite{reviebacte,predator,vegetation}.  
Different individuals struggle for nutrients 
with a competition strength directly dependent on the  
individuals' spatial density within the competing range.
Reproduction and/or death rates depending on the number of individuals 
in the surroundings can stand for this kind of interaction.
This feature has attracted the interest of experts from a variety of fields, ranging 
from pure mathematics \cite{math} and non-linear physics \cite{yosef,sasaki,kenkre,kenkre2,lopez1,continuous2} 
to population biology \cite{revew,popu}
and theoretical studies in evolutionary theory 
\cite{speciation1,speciation2,speciation3,speciation4,speciation5,speciation6}.
In addition, similar behaviors can be found in physical systems, such as, for example, in
mode interaction in crystallization fronts \cite{crystal} and in spin-wave patterns \cite{spin}. 
It is remarkable that the structured state generated by 
this kind of frequency-dependent interaction 
exists only for some specific 
form of the interaction \cite{speciation5} and is reached through 
a transition in the parameter space. 
This transition, ({\it segregation transition} \cite{yosef}), 
drives the steady state of the system from a spatially 
homogeneous distribution to one marked by some clearly 
distinguishable inhomogeneities.

All these models, characterized by diffusion effects and an implementation of 
frequency dependent birth and death processes, permit multiple interpretations.

In a common interpretation the system space directly 
represents the physical space 
where the organisms live and the diffusion represents their spatial movement. Competition
between individuals corresponds to   
a mechanism of growth control caused by 
limited common resources. In this case,
pattern formation can reproduce the evolution of 
bacterial colonies \cite{reviebacte}, plankton concentration \cite{nature}, 
development of vegetation \cite{vegetation}
or spatial distribution of  predators \cite{predator}.

On the other hand, a different interpretation 
enables us to describe the speciation process: 
the generation of two different species starting from 
one single continuous population of interbreeding organisms.
To be specific, we can describe the speciation process by representing all 
the phenotypic  characteristics that determine the biological success of an 
individual by a number, the strategy value, that labels each individual. 
By reproduction, that includes a mutation process, an
offspring inherits a strategy that slightly differs 
from that of its parent.
In order to model natural selection, a frequency-dependent mechanism 
that mimics competition, 
completes the ingredients necessary for the emergence of 
population clustering. In this scenario, the generation of a new cluster
is interpreted, in a broad sense, as a speciation event.
Now, if the model space represents the mentioned strategy space and 
the diffusion models the mutation process during reproduction, 
we can identify the mechanism of growth 
control with natural selection and the branching events 
with the speciation process. This different interpretation justifies 
the analogy between a model
that describes the speciation process and the ones that describe spatial 
pattern formation in the evolution of bacterial colonies, vegetation 
or predation.

We must remember that, since this model does not 
include sexual reproduction, we are describing trait divergence 
in an asexual population, rather than speciation. 
Anyway, apart from effects strictly related to sexual reproduction,
the dynamics characterized by the individuals' diffusion from regions of low viability in 
favor of more viable ones, is the essential core of these two phenomena.
A detailed and motivated discussion of these processes 
can be found in the following references 
\cite{speciation1,speciation2,speciation3,speciation4,speciation5,speciation6}.

We can describe such processes starting from an
individual based model, that yields information on the behavior of a finite system 
(finite population) and accounts for fluctuation effects caused by the discrete 
nature of the interacting agents. 
Another approach, that neglects fluctuations, describes individuals just with the use of a 
field that represents the population density at each position in space over time.
This method, usually denominated continuous mean-field description \cite{continuous1,continuous2}, 
becomes exact in the infinite-size limit if fluctuations are small compared 
to averages \cite{natureth3}.  
Note that we are not making a mean-field approximation in the nature 
of the interaction (see, for instance, ref.~\cite{sayama}), 
that, in our model, is local and different for each individual.

Choosing this second strategy, the generalization of a well-investigated equation  
(Fisher-Kolmogoroff-Petrovsky-Piscounoff \cite{FKPP}) 
is quite common at present. 
In addition to a diffusion process with coefficient $D$ and a population 
growth mechanism (rate $a$), this equation incorporates a growth-limiting 
process controlled by the
parameter $b$  \cite{sasaki,kenkre,lopez1}:

\begin{eqnarray}
\label{eq1}
\frac{\partial\rho(x,t)}{\partial t} =&D& \nabla^{2} \rho(x,t)+a\rho(x,t)-\\ \nonumber
&b&\rho(x,t)\int_{-\infty}^{+\infty} \rho(y,t)F(x,y) dy~,
\end{eqnarray} 
\noindent
where $\rho$ is the density of individuals at position $x$ and time $t$.
Competition is obtained by varying the death probability for each individual, and is 
controlled through the {\it influence function} $F(x,y)$. 
Let us focus on the shape of the influence function:
it can range from a simple box-like function to a globally uniform interaction.
However, the Gaussian function should be considered particularly
relevant.
If, for instance, we need to represent the activity 
of a sedentary animal the interaction represented in the influence 
function should take into account the individual's daily
excursion around the fixed breeding site that can be represented
by Brownian motion and, for this reason, by 
a Gaussian distribution.
In the same way, if we want to represent the habitat degeneration
induced by the growth  of a colony of plants \cite{plants} we can think that the colony 
is originated by a single individual that disperses its seeds in a way also 
well described by Brownian motion. 
More generally, for a biological interaction that does not 
stop at some  defined length (presence of a cutoff), and that is nonlocal 
and controlled by a purely stochastic process, the Gaussian function should be the 
most natural choice.
On the other hand, this choice is a source of complications. 
Deterministic descriptions, in the case of a Gaussian influence function, 
predict no pattern formation \cite{kenkre,polech}.
However, such descriptions do not take into account fluctuations arising 
from the discrete character of individuals.  
The importance of these fluctuations
has been recently pointed out in a quite paradigmatic example, 
where random walking organisms that reproduce and die at a constant 
rate spontaneously aggregate
\cite{natureth,natureth2,natureth3,nature,natureth1}.

The deterministic approximation is not able to show this behavior, 
incapable of capturing the essential asymmetry between birth, a multiplicative 
process that increments the density in the regions adjacent to the parent, 
and death events, that occur anywhere. 
Even when the patterns can be obtained within the 
deterministic description, a recent work outlines the importance of fluctuations 
by showing their impact on affecting transition points 
and amplitudes \cite{lopez1,lopez2,lopez3}.\\ 
 
In this work we present some numerical results obtained 
by means of a simple individual based model. 
Our aim is to show the appearance of
a  {\it segregation transition} in a model where the 
deterministic instability, produced by the non-local interaction, 
is not sufficient for generating inhomogeneities\cite{kenkre},
but the superimposed microscopic stochastic fluctuations
permit the emergence of patterns.
 Moreover, we compare  the model implementation in the strategy space with the implementation
in the physical space. In the first, used to characterize the speciation process, 
diffusion corresponds to a 
mutation phenomenon, operating just one time 
in each individual's life. 
In the second, directly related to the reaction-diffusion equation
(eq.~(\ref{eq1})), diffusion describes a typical Brownian motion.

The paper is organized as follows. The next section describes the model used in our simulations. Section III shows, for specific values of the parameters, the emergence of spatially inhomogeneous steady states. In section IV we prove that these patterns are not caused by some finite size effect. Section V is devoted to illustrate the segregation transition and general conditions that allow spatial segregation of arbitrary wavelengths. In section VI we describe, 
in the light of the existing literature, 
the cluster size dependence on diffusion rate and population size and give some hints related to the behavior of fluctuations. Conclusion are reported in section VII. 

\section{The model}

The simulations start with an initial population of $N_{0}$ 
individuals located along a ring of length $L$, i.e. we take periodic boundary conditions.
At each time step, our model is controlled by the following 
microscopic rules:

1) Each individual, characterized by its position $x$, dies with probability $P$,

\begin{eqnarray}
P = K \cdot \sum \limits_{j=1}^{N(\tau)} \exp(-\frac{(x-y_{j})^2}{2C^2})~,
\label{eq_death}
\end{eqnarray} 

where $N(\tau)$ is the total number of individuals at the actual 
step $\tau$ and $y_{j}$ is the respective individuals' positions.
The distance between two individuals is obtained by taking the shorter distance on
the ring. The strength of competition declines with increasing distance 
according to a Gaussian function with deviation 
$C$. The parameter $K$ depicts the carrying capacity.

2) If the individual survives this death selection,
it reproduces. The newborn, starting from the parent's location,  
moves in a random direction a distance obtained from a 
Gaussian distribution of standard deviation $\sigma$. 
This change represents the effect of mutations in the offspring phenotype.\\

As soon as all the individuals have passed 
the death selection and eventually reproduced, the next time step begins.
This model implementation is analogous to the 
diffusion process described by eq.~(\ref{eq1}).

To establish a more direct comparison between
that mean field description and the individual based 
simulation, we have also implemented the model with an
exact microscopic representation of the diffusion term.
In this case, at any given time step, we perform first a 
loop over all particles where 
individuals move some distance, 
in a random direction, chosen from a Gaussian 
distribution of standard deviation $\sigma$.
At the end of this loop, a second one starts,
where: 

1) Each individual with strategy $x$ dies with a probability 
obtained from eq.~(\ref{eq_death}).

2) If the individual survives the death selection process, 
it reproduces and the newborn maintains the same location 
of the parent.\\


If, in eq.~(\ref{eq1}), we measure time 
in units of the simulation time step, the coefficient 
$D$ is related to our simulation parameter 
through  $D \propto {\sigma^2}$. The influence 
function is given by a Gaussian of standard deviation 
$C$ and the effective growth rate is $1-P$, with $P$ given 
by eq.~(\ref{eq_death}). 
This non constant growth rate can be represented by eq.~(\ref{eq1}) 
for $a=1$ and the frequency dependent part included 
in the integral term (see ref.~\cite{lopez1}).


Essentially, the only difference between these two versions 
of our model is that, in the first one, individuals move only at 
birth, while in the second version they can move at every time step throughout their life.
Since the death probability, at equilibrium, is 
approximately $1/2$, usually, in the second version, 
an individual will move between one and two times during its entire life.
For this reason, there should be no relevant differences in the qualitative behavior of the
two model implementations. 
As shown by the measures reported in section IV, 
the only significant effect is the appearance of slightly wider distributions.

\section{Modulation}

 In the following we present some typical examples of 
steady states generated by the dynamics of the model that
clearly show the emergence of patterns 
for some specific values of the parameters.

\begin{figure}[p]
\begin{center}
\includegraphics[width=0.45\textwidth, angle=0]{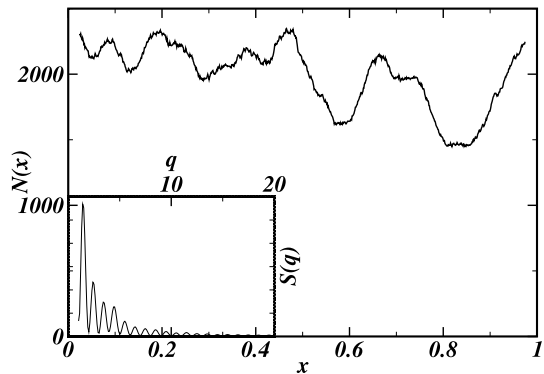}
\includegraphics[width=0.45\textwidth, angle=0]{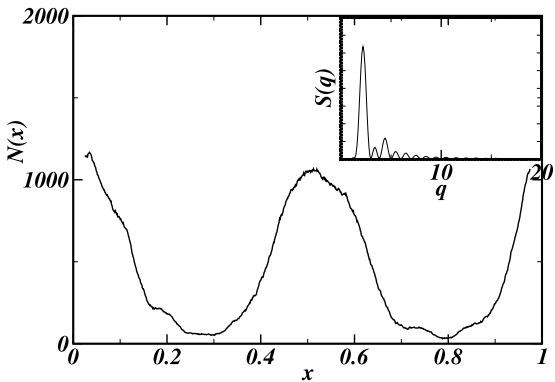}
\end{center}
\caption{\small Homogeneous steady state distribution ( top,
  $C=4.0,1/K=80000,\sigma=0.01$) and modulated steady state distribution (bottom,
  $C=0.9,1/K=18000,\sigma=0.01$). The insets show the structure functions,  
$S(q)$, of the corresponding simulations. We show the distributions at time step 1000,
  whereas the structure functions are averaged 
  over 500 time steps.  }
\label{fig_mod}
\end{figure}

For a global competition that results to be 
extremely long-ranged (large $C$ values, in relation to 
the values of  parameters $\sigma$ and $K$), the steady state is 
characterized by a spatially homogeneous occupancy. 
If the $\sigma$
value is sufficiently large or/and the $K$ value sufficiently small, 
totally homogeneous 
distributions are obtained  (Figure \ref{fig_mod}),
otherwise the solution is smooth but with 
the population concentrated in one region of the ring, 
with its width controlled by $\sigma$.\\

As stated above,  a simple heuristic analysis  of 
eq.~(\ref{eq1}) in the Fourier space shows that there is a necessary condition 
for the emergence of inhomogeneity: the Fourier transform of the 
influence function must have negative values and large enough 
magnitude \cite{kenkre,kenkre2}. A Gaussian in an infinite domain 
has a positive counterpart in Fourier space and so does not match 
such requirements.
In contrast with these results,
the fluctuations present in our individual based model 
arrive to excite one specific mode 
and modulations of this wavelength appear. 
The tuning of the parameters allows
modulations of arbitrary wavelengths (Figure \ref{fig_mod}).\\
 
\begin{figure}[p]
\begin{center}
\includegraphics[width=0.45\textwidth, angle=0]{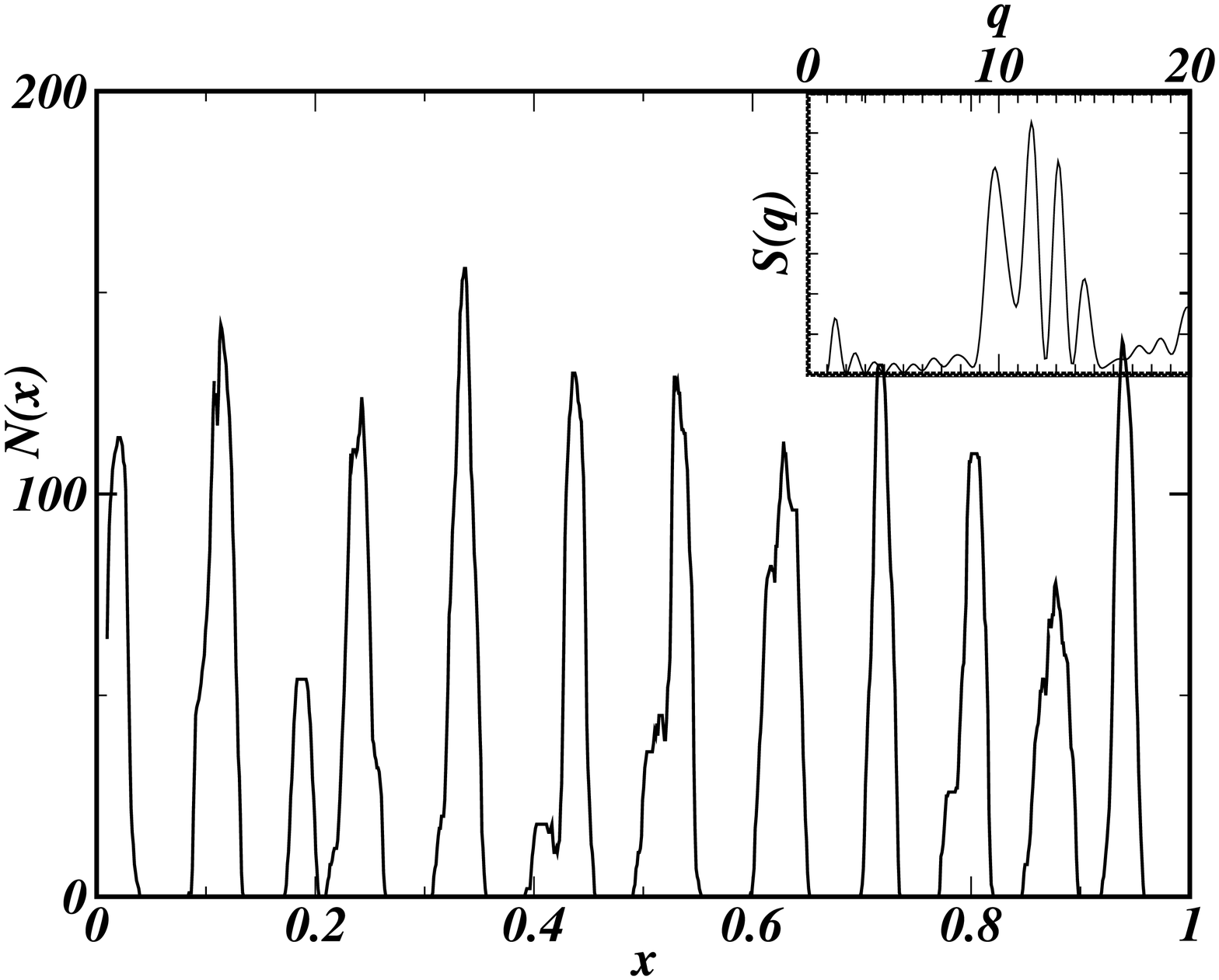}
\includegraphics[width=0.45\textwidth, angle=0]{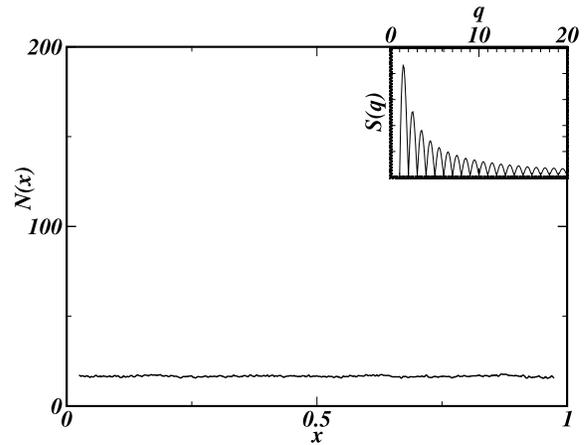}
\end{center}
\caption{\small Spiky (top, $C=0.059$, $1/K=500$, $\sigma=0.001$) and
homogeneous (bottom, $C=0.005$, $1/K=500$, $\sigma=0.001$)
steady states. 
The insets show the structure functions $S(q)$.   
We show the distributions at time step 2000,
whereas the structure functions are averaged
  over 1000 time steps. The transition between these two states, in
this typical range of parameters, has been extensively studied.} 
\label{fig_mod2}
\end{figure}

When $C$ is decreased, the competition between modes  
becomes stronger and no single mode dominates.  
In this situation, some small regions 
of the ring are occupied forcing all the remaining areas, up to some range, 
to be nearly empty.
The landscape results populated by several living colonies divided by dead regions. 
There is almost no competition between individuals of different colonies  
and the space separating them can be identified with an effective interaction length.
This steady state ({\it spiky state} \cite{yosef}) corresponds to  a sequence of isolated 
colonies (spikes) 
and seen in the Fourier space, many active wavelengths contribute to 
it (Figure \ref{fig_mod2}).\\

Finally, for extremely short-ranged competition, in relation to the $\sigma$ value, 
no collective cooperation  between different excited modes emerges 
and a noisy spatially homogeneous distribution appears (Figure \ref{fig_mod2}).\\ 

We describe these paradigmatic steady states  of the system by
characterizing the related spatial structure with the help of a 
{\it structure function} $S(q)$ \cite{lopez1} defined as follows:

\begin{eqnarray}
S(q) = \Big<\Big|\frac{1}{N(t)} \sum \limits_{j=1}^{N(t)} \exp[i 2\pi q\cdot x_{j}(\tau)]  \Big|^{2}\Big>_{T}
\label{stru}
\end{eqnarray} 

where the sum is performed over all individuals $j$
with their positions determined by $x_{j}(\tau)$. Note, that, for convenience, the
structure function is calculated only over the closed interval $[0,L]$.
The function is 
averaged over some time interval $T$ in order to avoid noisy data.
 $S(0)$ corresponds to  
the square of the mean number of individuals in the system. 
The maxima of $S(q)$ identify the relevant periodicity 
present in the steady state.
We will see that the position of the global maximum ($q_{M}$) 
provides an appropriate order parameter for the identification of 
the segregation transition.

In our study we explored two different initial conditions.  
In the first (local i.c.), 
the colony is located in a finite and compact region of the space. 
In the second  (global i.c.), the individuals are spread all over the space.
The final distribution is independent of this choice and,
generally, local initial conditions 
make the system reach the steady state more slowly. 
For this reason, if not differently specified, our results are  obtained 
from global initial conditions.

\section{Finite size effects}

Our analysis starts by exploring the model dependence on the 
space size $L$. 
 The reason for such interest is given by the necessity of 
testing if the pattern formation is not merely 
a product of some finite size effect. This is important,
in the light of what was reported by Fuentes {\it et al.} 
in ref. \cite{kenkre}. In their work, a numerical solution of 
eq.~(\ref{eq1}) with a Gaussian influence function, showed a segregation transition. 
But such a transition was just the effect of the finite 
domain size that acted like a cut-off for the Gaussian. 
 Evidence of this interpretation came from the observation that the amplitude 
of the pattern depended on the ratio of the standard deviation of the influence function to the domain size - the critical values of the standard deviation corresponding 
to the segregation transition depended linearly on the domain 
size - and the same patterns appeared for a modified Gaussian,
which vanishes abruptly beyond a cutoff. 

The study of our individual based model gave different results. By running some simulations 
with exactly the same parameters but changing the 
ring extension, we were able to show that the system is 
not influenced by the domain size. 
If we choose data from spiky steady states, that permit clear
quantitative measures, it is possible to  remark
that the general morphology of the patterns does not change 
increasing the $L$ value. 
In fact, both the population density and the mean number of peaks per space interval
  remains constant.  
Moreover, in
order to provide a more 
precise test of possible small variations in the distribution, 
we measured the cluster size. This quantity was calculated by evaluating the standard 
deviation $( <x^{2}>_{i}-{<x>_{i}}^{2})^{1/2}$ of the position of the $i$ individuals 
confined in each peak, then averaged over the different
peaks present at step $\tau$ and, finally, averaged over many time 
steps after the system has reached the steady state. 

Varying the system size caused 
no changes in the clusters size (see Figure \ref{fig_size}). 
From this result, we concluded that the general aspect of 
the steady state does not change with $L$. 
In particular, in contrast
with what happens when the mean field equation is solved numerically,
the patterns do not depend on the ratio $\frac{C}{L}$. 
For example, for  $C=0.2$ and $L=50$ we obtained a spiky steady state, for  $C=0.004$ and $L=1$  
(same ratio) we obtained an homogeneous steady state.
Taking into account  these results, from now on, all our simulations are 
implemented on a ring of size $1$.\\

We have just shown how the average of the population size $<N>$ 
scales with $L$ in the steady state. 
 Now, we will give, through a simple heuristic argument,
an estimation of $<N>$ as a function of the parameters $K$ and $C$,
that will be useful also in the following of our analysis.
 
We can assume that, locally and in the steady state, the number of deaths 
must be, on average, compensated by the number of newborns, in order to 
comprise a stable population. 
For this reason, the death probability $P$ must equal $1/2$. 
Assuming that the number of neighbors that compete 
with a single individual are the ones living up to a distance $C$ and that, 
in these surroundings, the average density $N/L$ can be considered to be uniform, 
$P$ reduces to $K\cdot 2C\cdot N/L$. Thus, $N \propto  L/(C K)$. 
Looking at Figure \ref{fig_pop} we can see that this crude 
evaluation, that neglects diffusion, inhomogeneity and reduces 
the influence function to a box, describes well the general behavior 
of the data obtained from our simulations.

\section{Segregation transition}
        
 In the previous paragraphs we supported the fact that steady states, 
depending on the parameter values, can assume inhomogeneous
spatial distributions. Now, we will try to describe the transition towards 
these states (segregation transition).
The structure function introduced in eq.~(\ref{stru}) provides a proper 
order parameter to describe this transition.
Different regions in the parameter space, coinciding with different 
steady states, correspond to different positions of the global maximum
(obviously we are not taking into account $S(q)$ at $q=0$)
of the structure function. 
The transition from 
a homogeneous to an inhomogeneous distribution (see Figure~2) matches the 
jump of the position of the global maximum ($q_{M}$) to a clear integer value, 
corresponding to the number of clusters present in the space.
For this reason we can characterize the transition by looking at 
the shape assumed by $S(q)$, or looking at the value of $q_{M}$. 
If the space is homogeneously occupied, the structure function does not 
present an integer maximum. On the contrary, the maximum is located 
at  $q_{M} \simeq 1.4$.
This value corresponds to an uniform distribution of individuals 
in the interval [0,1], approximated by the expression 
$\Big| \int_{0}^{1} \exp(i2\pi q\cdot x)dx  \Big|^2$. 
The segregation transition is characterized by the passage of $q_{M}$
from $1.4$ to an integer value as soon as a modulation becomes dominant. 
In Figure \ref{fig_trans} we show 
$q_{M}$ as a function of $C$, varying $K$ and $\sigma$. 
First of all, from the analysis of these data, we can observe that 
the number of clusters scales as $C^{-1}$ (or, equivalently, the 
periodicity of the inhomogeneous phase has wave lengths proportional to $C$). 
Moreover, a critical value of
$C$ exists for which the transition takes place. This $C_{critic}$ grows
with $1/K$ and with $\sigma$. An analysis of the available data suggests the possible dependence: 
$C_{critic} \propto \sigma^{2/3} K^{-1/3}$ . 
Finally, for larger values of the parameter $C$,
in this range of the parameters $\sigma$ and $K$, 
the distributions are characterized by just one peak. \\

For any value of the competition strength, as can be seen in 
Figure \ref{fig_difcrit}, there exists a critical value  
$\sigma_{critic}$, dependent on $C$ and $K$, 
above which no spatial structures emerge. 
Another measurement, that permits us to state this relation in a 
different and clearer way, is presented in the next paragraph.

\section{Cluster size and fluctuations}

 In the following we describe, in the light of the existing literature, the
cluster size dependence on diffusion rate and population size for the two 
different implementations of the model.
 
We start by analyzing the typical size $S$ of the clusters that appear in
the spiky phase.
The data exposed in Figure \ref{fig_clus} show a dependence of 
the cluster size on the diffusion coefficient: $S \propto \sigma$, equivalent to 
$S \propto \sqrt{D}$. 
These results are in accordance with the data presented in ref.
\cite{lopez3} obtained from an individual based model. 
In addition, this work pointed out how this behavior deviates 
from the conclusions obtained from the deterministic approximation,
where the cluster size was only weakly dependent on the diffusion coefficient; 
 another fact supporting the relevance of fluctuation effects in these systems.

We can easily interpret the dependence of the cluster size on the diffusion coefficient by
assuming that the individuals confined in a cluster 
diffuse a distance proportional to  $\sqrt{DJ}$ where $J$ is the 
number of jumps the individual performs in its life. In the case of the 
first implementation of the model, where the diffusion is due to the 
mutation process, $J$ is obviously $1$.  
Similar results are obtained  
with the second implementation (see Figure~\ref{fig_clus}), 
apart from a slightly wider cluster size (in this
case, individuals move, on average, 
more than just once during their lifetime).
Even so, the data show the same dependence on the 
diffusion coefficient.

Finally, we present $S$ as a function of $K$: 
$S\propto \sqrt{1/K} $.
The reasons for this behavior are already explained in ref. \cite{lopez4}:
the cluster size is not controlled just by the single individual's
number of jumps; in fact the diffusive process 
continues with its descendants.
For this reason, it is proportional to the mean 
lifetime of a family, estimated to be proportional to $K^{-1}$ 
(see ref. \cite{lopez4} for details).\\

 We introduce a new quantity that is useful for 
describing the existence of a critical diffusion value, giving an 
estimation of its dependence on others parameters and 
a confirmation of previous results.
This quantity, which we call  {\it mobility} $M(\tau)$, estimates the mean mobility of individuals.
At a given time step $\tau$, we choose an individual $i$.
Then we look for the closest agent, among all the population, 
at time step $\tau-1$.  
We identify as  $d_{i}$ the distance between these two individuals.
Averaging over the entire population $N(\tau)$ we obtain:
\begin{eqnarray}
M(\tau) = \frac{1}{N(\tau)} \sum \limits_{i=1}^{N(\tau)}d_{i}.
\label{mob}
\end{eqnarray}

The values assumed by $M$ on varying the parameters $\sigma$ and $C$ 
are shown in Figure \ref{fig_mob}. It is easy to distinguish two
clearly different behaviors. If the system is organized in a spiky 
state (when $\sigma\le\sigma_{critic}$), $M(\tau) \propto \sigma$. $M$ is another way of measuring 
the mean distance that an individual moves during its lifetime inside
the region defined by the cluster. 
For this reason, this measure is coherent with the data obtained 
from the direct 
evaluation of the standard deviation of the clusters. 
In contrast, when the system is organized in the homogeneous 
phase (when $\sigma>\sigma_{critic}$) $M$ becomes independent of $\sigma$ 
and is proportional to the inverse of the occupation density 
$M(\tau) \propto KC$. 
The values of  the mobility obtained from simulations with different values of 
$C$ and $K$ can be easily collapsed into one function (see the inset of Figure \ref{fig_mob}). 
The collapse is performed using the scaling $\sigma \to \sigma/C\sqrt{K}$.
This indicates that the characteristic value of the  
crossover, $\sigma_{critic}$, that separates the two different behaviors of $M(\tau)$, scales as: 
\begin{eqnarray} 
\sigma_{critic}\propto C\sqrt{K}.
\label{dcritic} 
\end{eqnarray}\\ 

We conclude our study with some measurements trying to catch 
some properties of the system fluctuations. First, we estimated the fluctuations of the total population, averaging 
over different simulations. The variance turned out to be 
constant throughout the time evolution and 
of the order of the square root of the total population. 
The mechanism of 
auto-regulation of the population dimension
does not allow the growth of big differences in the total number
of individuals. 

For this reason, we focused our attention on the 
spatial distribution of the population 
and tried to measure some properties of these fluctuations.
We studied the variation of the local number of individuals
in the same simulation, for different times. 
We analyzed the evolution of 
the system starting from local initial conditions, with the population 
concentrated in the interval $[0.49,0.51]$. 
In this situation, the system evolves in time with a small cluster
fluctuating around the initial space interval. This situation changes 
when a branching event occurs that generates two well-defined clusters. 
Our interest is in showing the behavior of local space fluctuations 
and capturing possible variations in correspondence with the branching 
event. First of all, we looked at the mean value of the spatial local 
fluctuations  $F_{s}(\tau)$, defined as 
$F_{s}(\tau) =(\sum \limits_{j=1}^{b}f_{j}^{2})^{1/2}$,  
where $f_{j}$ is the occupancy variation of the bin $j$ 
from time step $\tau-1$ to time step $\tau$. We performed the 
average over all  the $b$ bins the ring was divided in
and obtained $F_{s}(\tau) = \sqrt{N(\tau)}$,
with no relevant variations throughout the time evolution,
even in the time interval corresponding to the branching event. 
More interesting is the shape of the frequency distribution of the size of
$f_{j}$. 
In fact, a simple Gaussian does not fit this distribution, 
that presents extended tails (see Figure \ref{fig_fluc}). 
Throughout the system time evolution, the shape of the normalized distribution is conserved. 
For global initial conditions the same  
frequency distribution, with extended tails, is recovered
at the steady state. 
It is identical to the one obtained with local initial conditions 
and measured at the steady state.
We think that  the deviation of the  distribution from a Gaussian can be considered 
as a hint that fluctuations play a relevant role in the dynamics of the systems.

\section{Conclusions}

We presented some results regarding clustering of organisms 
caused by a frequency-dependent interaction that represents competition. 
We showed 
how this way of modeling competition can be used 
not only to describe spatial phenomena in population 
biology, but also, through a more abstract interpretation,
to test ideas of evolutionary theory 
(for example, studying the speciation process).

From this unifying perspective, our study, obtained from an 
extensive collection of data coming from simulations of an 
individual based model with global competition,  
pointed out the relevance of  fluctuation effects in pattern 
formation. For the influence function adopted, 
the mean field description predicts the absence 
of spatial structures. On the contrary, fluctuations are able 
to excite the emergence of well defined patterns, which can 
not be generated from a deterministic instability.

Furthermore, we discussed other fundamental properties of our model
in the light of the existing literature, unfolding a comparison with other models 
that describe spatial segregation originated by some 
deterministic instability.
We showed that the observed patterns are not due to a finite size effect, we characterized 
the behavior of the segregation transition in various regions of 
the parameter space and we studied the existence of a critical diffusion value.
We analyzed the dependence of the cluster size on the diffusion 
coefficient and pointed out some characteristics of the fluctuations 
of the system. 

\section*{Acknowledgments} 

We are grateful to J.S. S\'a Martins for a critical reading of the manuscript and thank
the Brazilian agency CNPq for financial support.

\begin{figure}[p]
\begin{center}
\includegraphics[width=0.45\textwidth, angle=0]{size.eps}
\end{center}
\caption{\small Dependence of the cluster size on the ring size $L$. 
We present data from the model with mutation (triangles) and
from the one that implements diffusion (circles), $C=0.2,K=0.0029,\sigma=0.001$. 
The average is carried out over all the clusters
present at a given time step and over different time steps.} 
\label{fig_size}
\end{figure}

\begin{figure}[p]
\begin{center}
\includegraphics[width=0.45\textwidth, angle=0]{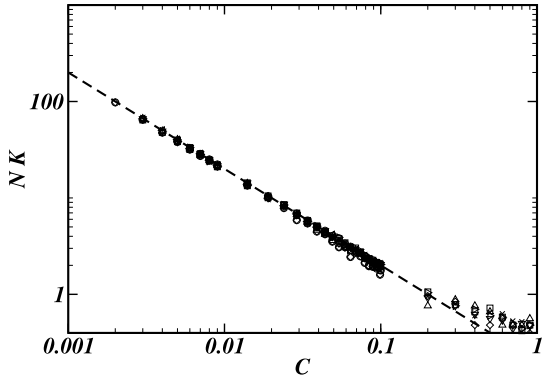}
\end{center}
\caption{\small The number of individuals $N$ present in the steady state is
proportional to $(C K)^{-1}$. This result is in accordance with the one
obtained for a box-type influence function of length $C$ (see ref.~\protect\cite{lopez2}). 
We present data for different simulations with $1/K \in [50,500]$ and
$\sigma\in [0.0001,0.01]$. 
This last parameter does not influence the final number of individuals. The dashed
line has slope -1.}
\label{fig_pop}
\end{figure}

\begin{figure}[p]
\begin{center}
\vspace*{0.8cm}
\includegraphics[width=0.45\textwidth, angle=0]{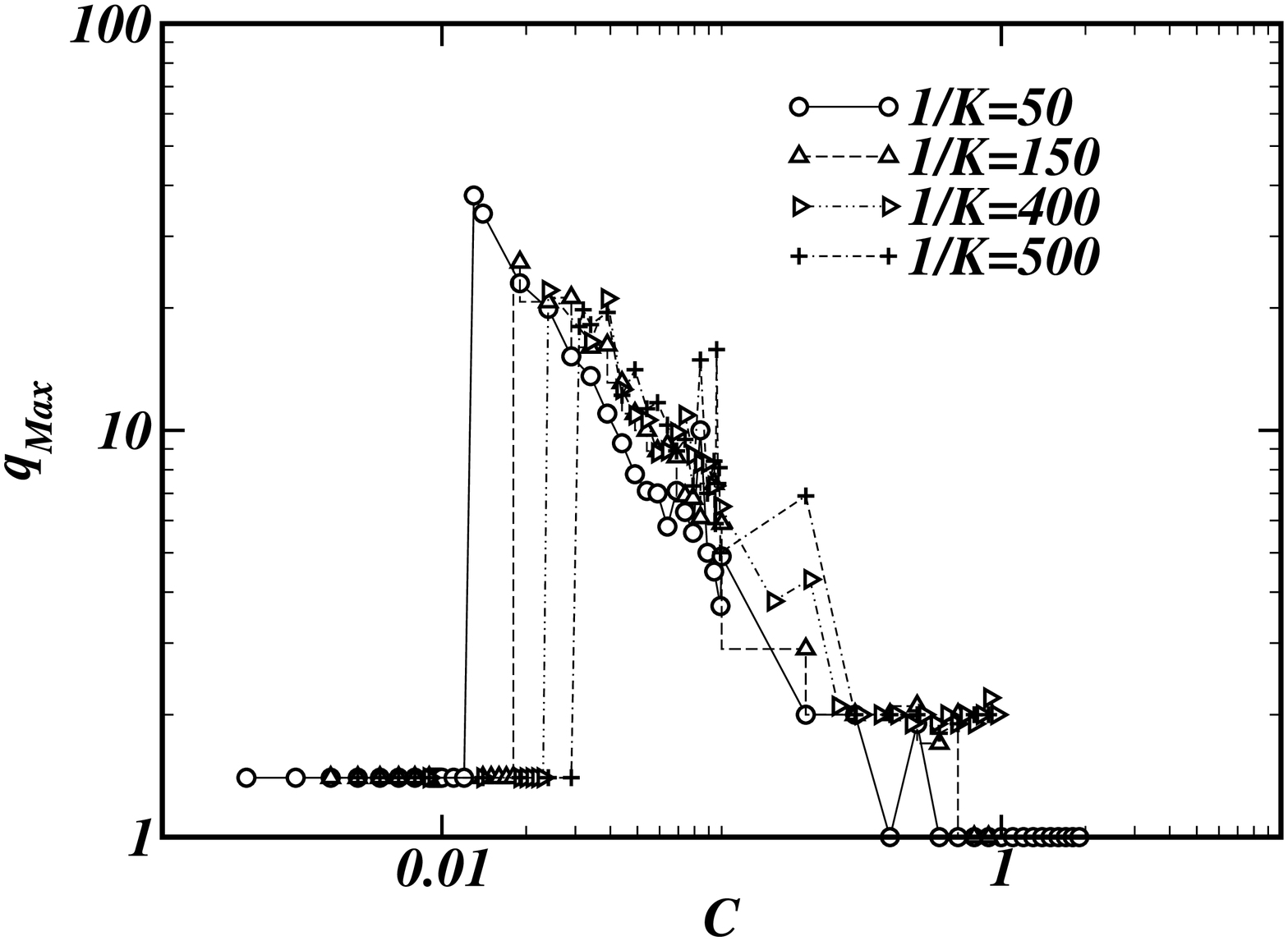}
\includegraphics[width=0.45\textwidth, angle=0]{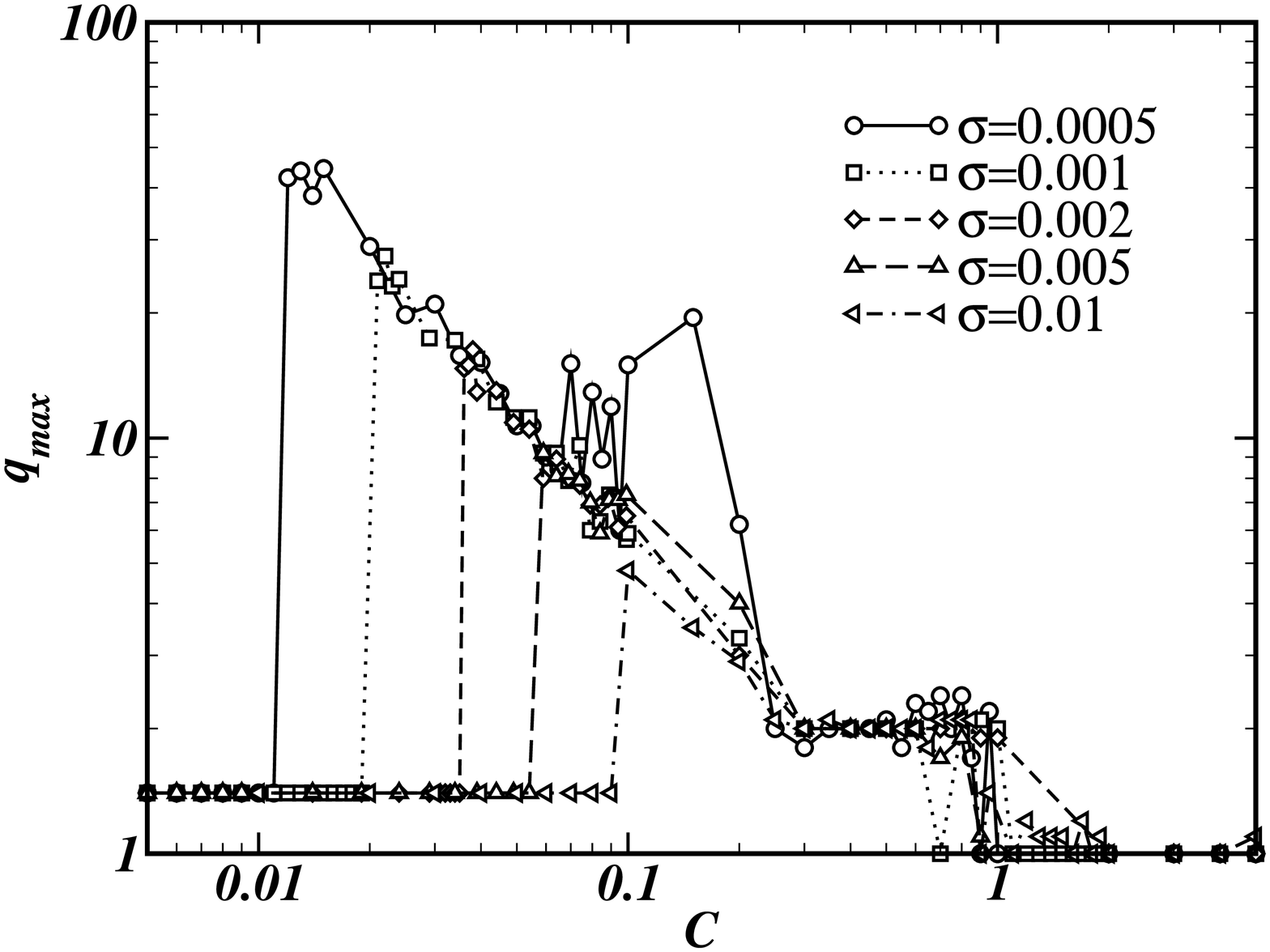}
\vspace*{0.4cm}
\end{center}
\caption{\small Segregation transition at $C_{critical}$. 
Upper figure: variation in dependence of $K$, where $1/K=50,150,400,500$ and $\sigma=0.001$. 
Lower figure: variation in
dependence of $\sigma$, where $\sigma=0.0005$, $0.001$, $0.002$, $0.005$, $0.01$ and $1/K=200$.}
\label{fig_trans}
\end{figure}

\begin{figure}[p]
\begin{center}
\includegraphics[width=0.45\textwidth, angle=0]{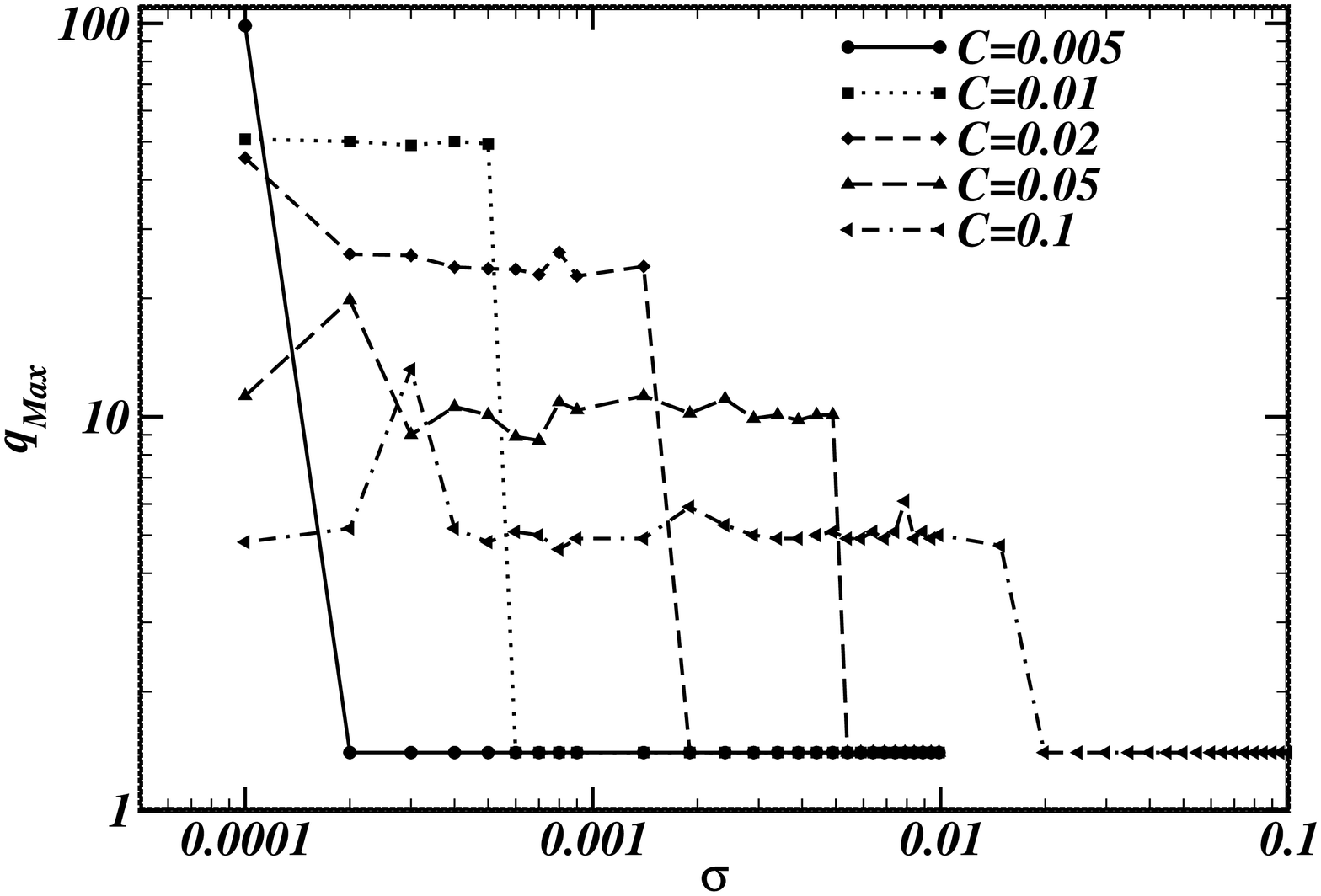}
\includegraphics[width=0.45\textwidth, angle=0]{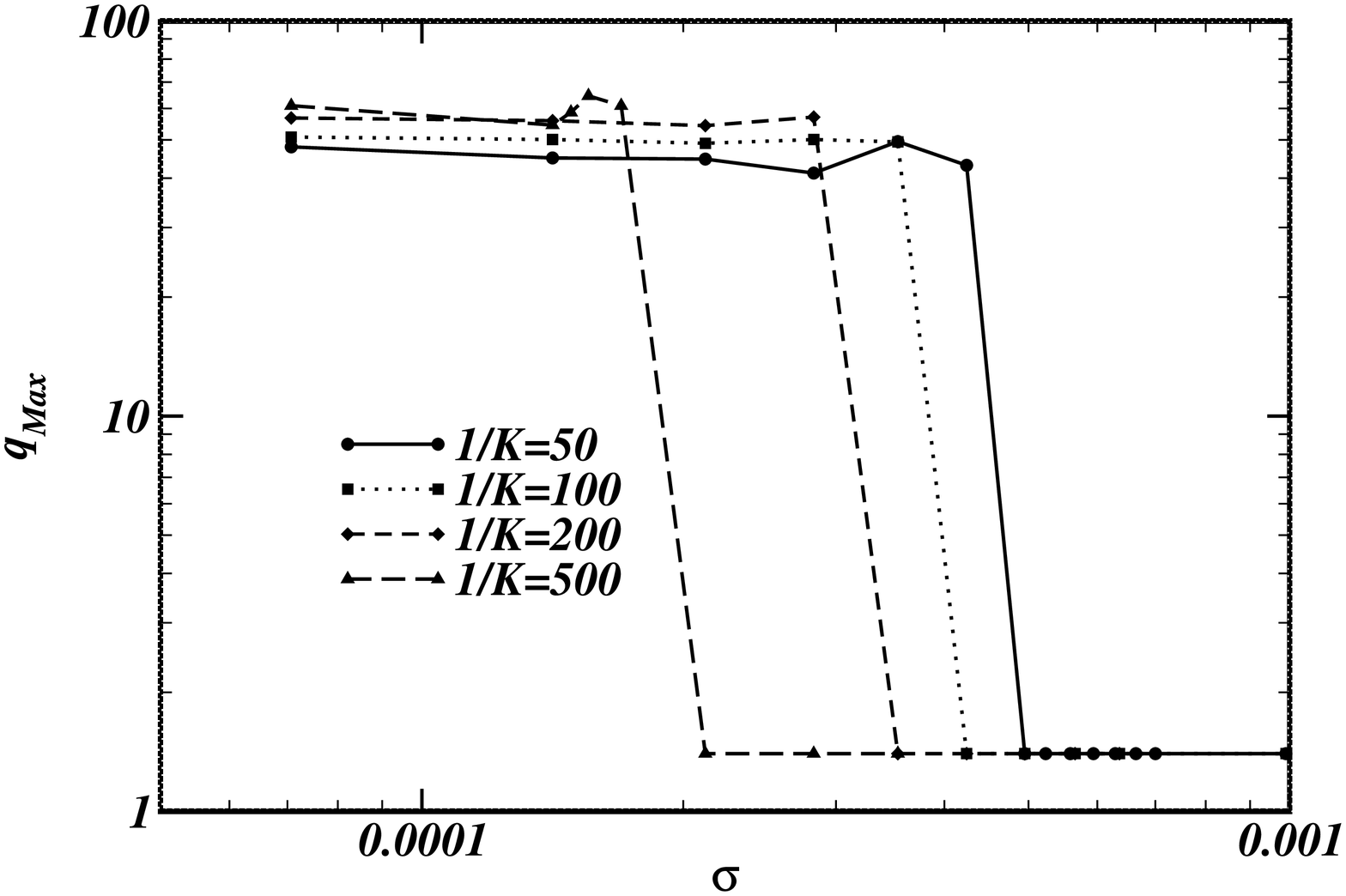}
\end{center}
\caption{\small As shown in these figures, a critical 
$\sigma$ value exists, above which no spatial structures emerge. 
Upper figure: variation in dependence on $C$; we set $1/K=100$. Lower
  figure: variation in dependence on $K$; we set $C=0.01$.}
\label{fig_difcrit}
\end{figure}

\begin{figure}[p]
\begin{center}
\includegraphics[width=0.45\textwidth, angle=0]{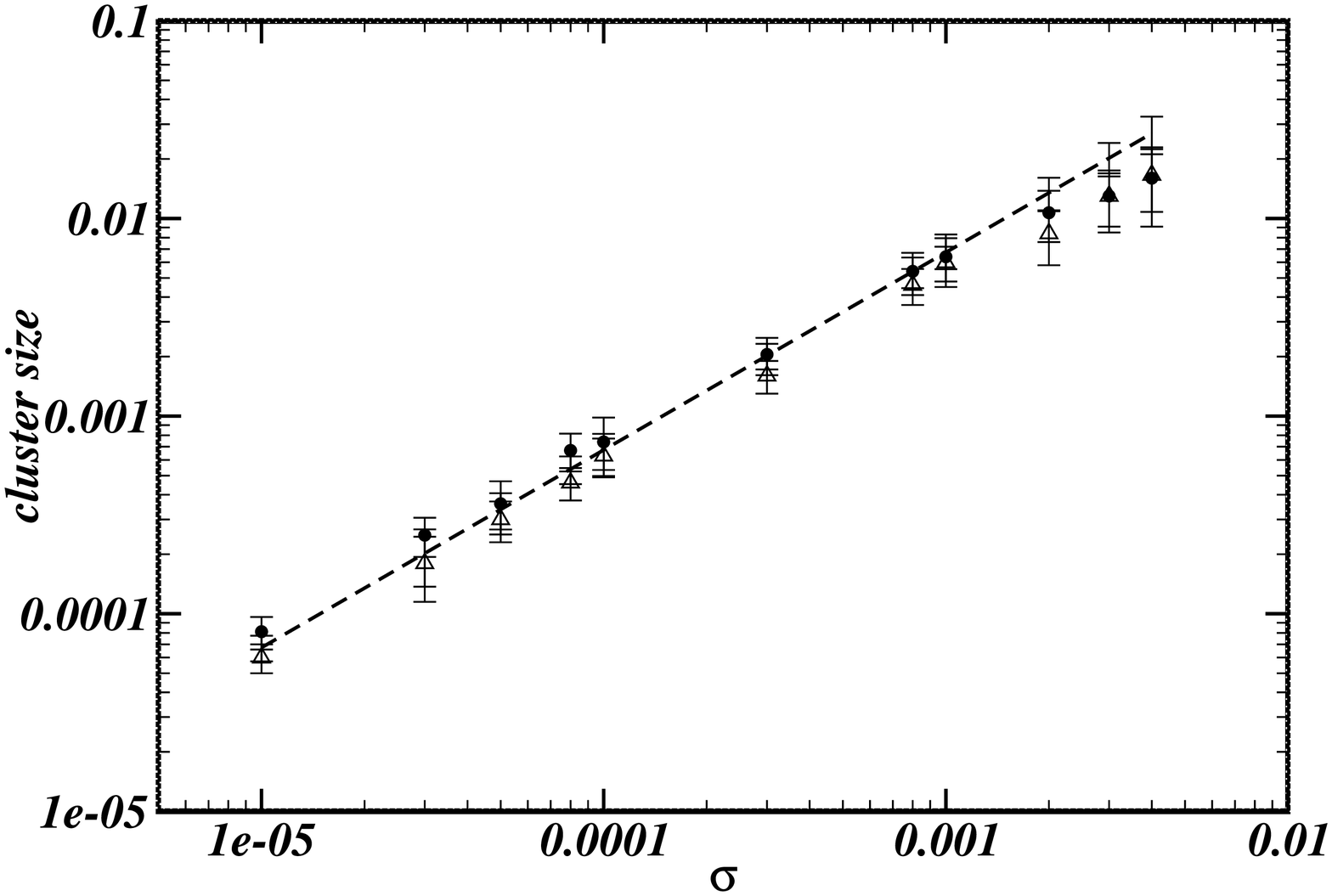}
\includegraphics[width=0.45\textwidth, angle=0]{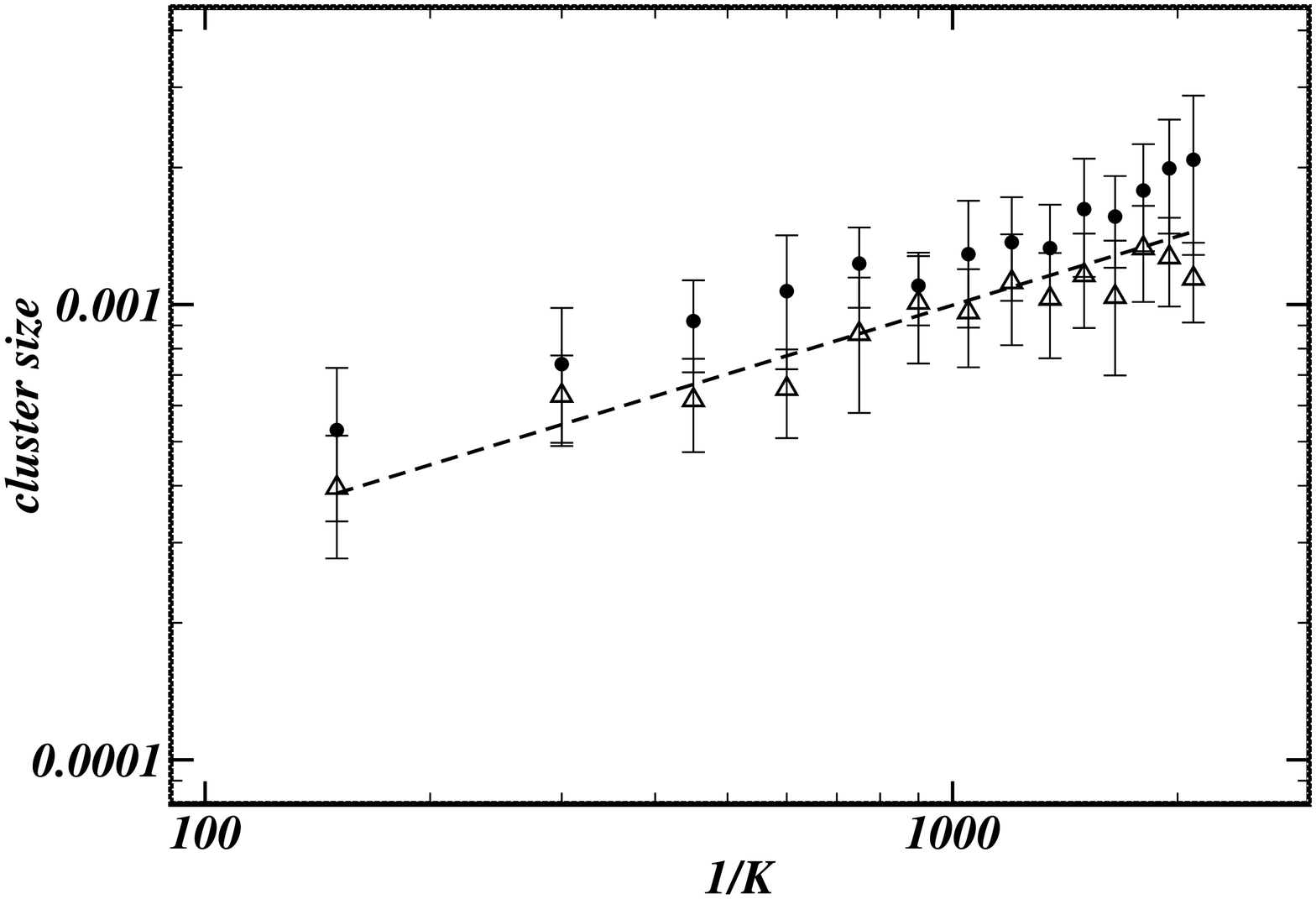}
\end{center}
\caption{\small Top: Cluster size as a function of $\sigma$; $1/K=300$. 
The solid line has slope 1. 
Bottom: cluster size as a function of $1/K$; $\sigma=0.0001$. 
The solid line has slope 1/2.  
 Triangles represent data from the simulations where
diffusion is implemented through mutations, circles for the direct
implementation of the diffusive process; we set $C=0.09$.}
\label{fig_clus}
\end{figure}

\begin{figure}[p]
\begin{center}
\includegraphics[width=0.45\textwidth, angle=0]{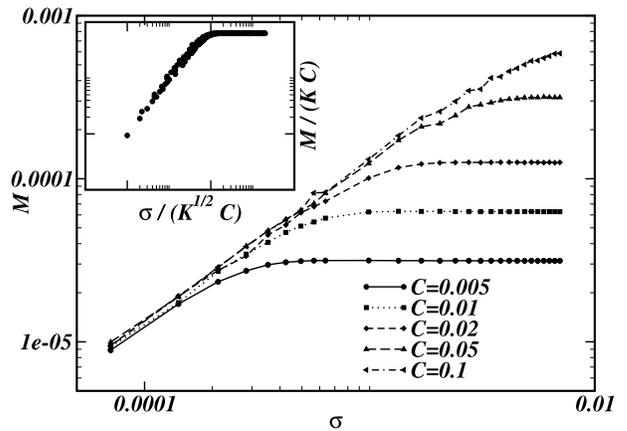}
\end{center}
\caption{\small Mobility dependence on the diffusion parameter 
$\sigma$ for different $C$ values; $K=0.01$.  In the inset, the data collapse for arbitrary
values of the parameters $C$ and $K$.} 
\label{fig_mob}
\end{figure}

\begin{figure}[p]
\begin{center}
\vspace*{0.8cm}
\includegraphics[width=0.45\textwidth, angle=0]{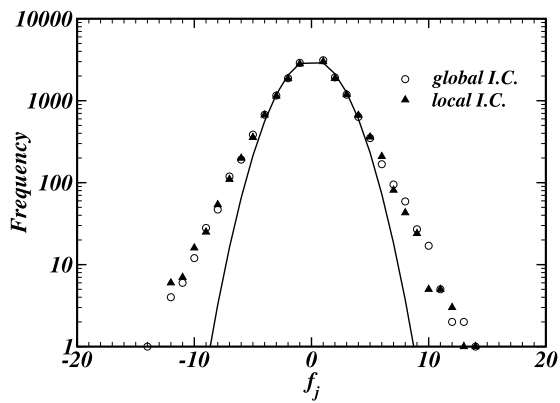}
\vspace*{0.4cm}
\end{center}
\caption{\small Spatial local fluctuation distribution at the steady 
state: deviation from a Gaussian 
($C=0.1$, $K=0.00005$, $\sigma=0.0017$). 
Data are averaged over $5$ time steps. 
The continuous line is the best fit Gaussian.}
\label{fig_fluc}
\end{figure}

\end{multicols}

\end{document}